           \documentstyle[prl,preprint,aps]{revtex}% <--|  <----| <--|

\begin{document}

% Here comes the title page ***************************************************
%\draft
\title{
                     The Earth Effect in the MSW Analysis  \\
                     of the Solar Neutrino Experiments
}
\author{
                     Naoya Hata and Paul Langacker
\\
}
\address{                                                         %**revtex**
                     Department of Physics,                       %**revtex**
                     University of Pennsylvania,                  %**revtex**
                     Philadelphia, PA 19104                       %**revtex**
}                                                                 %**revtex**
\date{
                     May 3, 1993, UPR-0570T %, Draft Version 1.1
}
\maketitle

% Here comes the abstract *****************************************************

\begin{abstract}

We consider the Earth effect in the MSW analysis of the Homestake, Kamiokande,
GALLEX, and SAGE solar neutrino experiments.  Using the time-averaged data
and assuming two-flavor oscillations, the large-angle region of the combined
fit extends to much smaller angles (to $\sin^22\theta \simeq 0.1$) than when
the Earth effect is ignored.  However, the additional constraint from the
Kamiokande II day-night data excludes most of the parameter space sensitive
to the Earth effect independent of astrophysical uncertainties, and leaves
only a small large-angle region close to  maximal mixing at 90\% C.L.
The nonadiabatic solution remains unaffected by the Earth effect and is still
preferred.  Both theoretical and experimental uncertainties are included in
the analysis.

\end{abstract}
\pacs{PACS numbers: 14.60.Gh, 96.60.Kx}                           %**revtex**

%\twocolumn      %**twocolumn**
\newpage

% Here comes the text *********************************************************

The recent results from SAGE, GALLEX, and Kamiokande III  confirmed the
deficit of the neutrino flux from the Sun, and thus all existing solar neutrino
experiments report a discrepancy at some level between the neutrino flux
observed and predicted \cite{Homestake,KII,KIII,SAGE,GALLEX}.  If the
experiments are correct, the
implications are significant: first the standard solar model is excluded
by the data \cite{Bahcall-Bethe,Langacker93}.
Secondly, the smaller chlorine rate relative to the
Kamiokande rate makes astrophysical explanations difficult and
excludes a wide class of nonstandard
solar models described by a lower core temperature
\cite{BHKL,BKL,GALLEX}.
On the other hand, among many proposed particle physics solutions,
the Mikheyev-Smirnov-Wolfenstein (MSW) mechanism \cite{MSW}
remains as a viable description of the data, giving a strong hint for
neutrino mass.

To explore this possibility, it is important to determine
the MSW parameter space from the available data.
The MSW effect
is an experimentally verifiable hypothesis;
once the parameter space is constrained, one can provide  robust
predictions for the next-generation detectors such as SNO, Super-Kamiokande,
BOREXINO, and ICARUS.   Theoretically, the
neutrino mass and mixing angle, along with the seesaw mechanism,
can be a probe of the physics at higher energy scales
({\em e.g.}, the grand unification scale) inaccessible
to laboratory experiments.

If we assume two-flavor MSW oscillations into $\nu_\mu$ or $\nu_\tau$,
the combined data of all experiments allow two solutions
\footnote{
There is also a nonadiabatic solution for oscillations into
sterile neutrinos, which is not affected significantly by the
Earth effect.  There is no large-angle solution for sterile
neutrinos at 90\% C.L. \cite{BHKL}.}
 with the squared mass difference $\Delta m^2 \sim 10^{-5}\mbox{eV}^2$:
one in the nonadiabatic region with mixing angle $\sin^22\theta \sim 0.01$ and
the other in the large-angle region with $\sin^22\theta \sim 0.8$,  the former
solution giving the better fit
\cite{BHKL} (see also
\cite{GALLEX,Shi-Schramm,Gelb-Kwong-Rosen,Krastev-Petcov,Krauss-Gates-White}).
There is one complication in the large-angle MSW solution.  For the
 $^8\mbox{B}$ neutrino
energy range, the MSW parameters satisfy the resonance condition for the
electron density  corresponding to the core and mantle of the
Earth.  The $\nu_x$'s, into
which electron neutrinos convert in the Sun, can
oscillate back to electron neutrinos when going through the Earth.
In such a case, the neutrino signals  should be enhanced during the nighttime
and can change the allowed parameter space of the combined fit.
In this paper, by including the Earth effect, we improve the global analysis
of  Bludman, Hata, Kennedy, and Langacker, who incorporated
theoretical and experimental uncertainties in a $\chi^2$
MSW analysis \cite{BHKL}.

Neutrino oscillations in the Earth have been discussed in detail by a number
of authors
\cite{Earth-effect}.  The survival probability $P_s$ of an electron neutrino
reaching the Earth from the core of the Sun where it was created
is calculated as
in Ref.\cite{BHKL}; the phase difference of the two neutrino mass
eigenstates at the
entrance of the Earth can be  averaged when finite energy bins and source
broadening are included \cite{Earth-effect}.  Then
the $\nu_e$ survival probability detected after propagating through
the Earth is
$$
P(\nu_e \rightarrow \nu_e) = P_s |a|^2 + (1-P_s) |b|^2
                          - (P_s - \frac{1}{2}) \tan 2\theta (ab^* + a^*b),
$$
where  $a$ and $b$ are the components of the unitary matrix which describes
the Earth effect:
$$
\left( \begin{array}{c}

	\nu_e (t) \\
	\nu_x (t)

       \end{array} \right)_{detector}  =
\left( \begin{array}{cc}

	 a    &  b    \\
	-b^*  &  a^*

       \end{array} \right) \left( \begin{array}{c}

					\nu_e(0) \\
					\nu_x(0)

				  \end{array} \right)_{surface}
$$
They depend on the energy,
$\Delta m^2$, $\sin^22\theta$, the local electron density, and the path length.
In Refs.\cite{Earth-effect}, the transition matrix is calculated by
solving the propagation equation numerically.  Here, however,
we obtain an analytic formula  by approximating the density profile with
a series of either
five step functions or five linear functions with respect to the Earth
radius \cite{Earth-density}.
This sudden approximation is justified since the dominant effect of
neutrino oscillations in the Earth is the matter oscillation length becoming
comparable to the neutrino path length, rather than a level crossing
between the two energy states as in the oscillations in the Sun.
Our calculations are in agreement  with those
of Baltz and Weneser and
of Carlson \cite{Earth-effect}.

The survival probability depends sensitively on the neutrino path
length and thus varies with time, date, and detector latitude.
If the parameters are in the range
$\Delta m^2/E = 10^{-7}-10^{-6} \mbox{eV}^2/\mbox{MeV}$
and
$\sin^22\theta > 10^{-2}$,
the solar neutrino signals should show not only different counting
rates between day and night, but also a distortion of the energy spectrum
with time, and seasonal variations during nighttime, which provides an
opportunity for
direct-counting detectors to constrain  the parameter space precisely.
To compare with the time-averaged experimental results,
we integrate the survival probability
over each night and over a year, taking about thirty points for each.

We use two data sets to constrain $\Delta m^2$ and $\sin^22\theta$.  One is the
time-averaged
rates of the Homestake, the combined Kamiokande II and III, and the gallium
results from SAGE and GALLEX
\footnote{In principle the two predicted gallium rates differ  because of their
detector latitudes. However,  the experimental uncertainties are
still large and the difference is insignificant.  We combine the two results
and use 42.4$^\circ$N as the latitude. }
, which are listed in
Table~\ref{experiments}.  The additional information from the Kamiokande
day-night data further constrains the parameters.  We use the
Bahcall-Pinsonneault solar model including the helium diffusion effect
\cite{Bahcall-Pinsonneault}.

In fitting the data by the $\chi^2$ procedure, the
experimental errors as well as all theoretical
uncertainties due to the initial flux and the detector cross
sections are incorporated.
The $\chi^2$ value for the day-night averaged data
 is calculated for
each $\Delta m^2$ and $\sin^22\theta$ by
$$
\chi^2 (\Delta m^2, \sin^22\theta) = \sum_{i,j = Kam, Cl, Ga}
(V^{-1})_{ij}(R_i^{exp} - R_i^{MSW})(R_j^{exp} - R_j^{MSW}),
$$
where $R_i^{exp}$ and $R_i^{MSW}$ are the rates of the experiments
and the MSW predictions for detector $i$ (= Kamiokande, Cl, and Ga), and
$V$ is the $3\times3$ error matrix.  The diagonal
elements are quadratic sums of the uncertainties:
the experimental errors, detector cross section
uncertainties ($\Delta \sigma_i/\sigma_i=$3.3\% and 4\% for the
chlorine and gallium detectors, respectively
\cite{Bahcall-Ulrich}), and
 the flux uncertainties according to the
Bahcall-Pinsonneault model \cite{Bahcall-Pinsonneault};
$$
    V_{ii} = (\Delta R_i^{exp})^2 +
             \left(\frac{\Delta \sigma_i}{\sigma_i} R_i^{MSW}\right)^2 +
	     \sum_{n=pp,^7Be,^8B, \cdots}
            \left(\frac{\Delta \phi_n}{\phi_n} R_i^{n,MSW}\right)^2,
$$
where $R_i^{n,MSW}$ is the MSW predictions for the $n$-th flux $\phi_n$
($n$ = $pp$, $^7\mbox{Be}$, $^8\mbox{B}$, $pep$, $hep$ and CNO).
The theoretical flux uncertainty $\Delta \phi_n$ includes contributions
 from  the uncertainty in the core temperature and in the
production cross sections, added quadratically.
The off-diagonal elements describe
the correlations of the flux uncertainties among the experiments;
$$
    V_{ij} = \sum_{n=pp,^7Be,^8B, \cdots}
            \left(\frac{\Delta \phi_n}{\phi_n}\right)^2
             R_i^{n,MSW}R_j^{n,MSW}
            \hspace{2em}  (i \neq j).
$$
We emphasize the importance of those correlations since the dominant
theoretical uncertainty from the $^8\mbox{B}$ flux (14\% at $1\sigma$) is
strongly correlated between the Kamiokande rate and the chlorine rate,
and also because the uncertainties in the different flux components are
correlated.   Ignoring the correlation, we obtained significantly larger
allowed regions, especially for the large angle solution
\footnote{The authors of  Ref.\cite{Krauss-Gates-White} show larger
allowed  MSW regions compared to ours, because of the
omission of  the correlations of the flux uncertainties.  Moreover,
the uncertainties used in Ref.\cite{Krauss-Gates-White} are
larger than the estimations of Bahcall-Pinsonneault.}.

The confidence level (C.L.) contours in  the
 $\Delta m^2$ and $\sin^22\theta$ space are defined by
$$
  \chi^2(\Delta m^2, \sin^22\theta) = \chi^2_{min} + \Delta \chi^2,
$$
where $\chi^2_{min}$ is the global minimum of the $\chi^2$ value in the
entire
$\Delta m^2 - \sin^22\theta$
plane and  $\Delta \chi^2 =$ 4.6, 6.0, and 9.2 for 90, 95, and 99\% C.L.,
respectively.  Caution is in order in determining the confidence
level contours.
The $\chi^2$ procedure assumes Gaussian distributions for the two
parameters, which is only an approximation in the present  case.
Furthermore the statistical interpretation is not clear when there is more
than one allowed region.  Therefore the confidence level contours
should be considered as qualitative.

The combined fit of the time-averaged Homestake, Kamiokande II and III, SAGE,
and GALLEX
results are displayed in Fig.~\ref{fig_ave}.  For the large-angle solution the
$\nu_e$ regeneration in
the Earth for the $^8\mbox{B}$ neutrinos enhances the nighttime rates of  the
Kamiokande and Homestake experiments for
$\Delta m^2 = 10^{-6}-10^{-5} \mbox{eV}^2$,
and enlarges the allowed region of the combined
fit down to
$\sin^22\theta \simeq 0.1$ at 90\% C.L.  In particular the Cabibbo-like
solution
$\sin^2 2\theta_{e\mu} = 0.18$
is now possible for
$\Delta m^2 \simeq 4\times 10^{-6} \mbox{eV}^2$.

An additional constraint comes from the Kamiokande II experiment,
which presented the results with six bins according to the
angle $\delta_{sun}$ between the vector pointing from the detector to the Sun
and the nadir at the detector \cite{KII-day-night}.
The first bin ($\cos \delta_{sun} < 0$) is
the daytime rate; the nighttime rates are for
$\cos\delta_{sun} =$ 0 -- 0.2, 0.2 -- 0.4, 0.4 -- 0.6, 0.6 -- 0.8, and
0.8 -- 1.0, which correspond to binning the neutrinos with different
 path lengths in the Earth.

The Kamiokande II day-night data \cite{KII-day-night} are consistent with
no enhancement of the
signal during the night and exclude the parameter space shown in
Fig.\ \ref{fig_dn} \cite{KII-day-night}, which is in good agreement with the
excluded region presented in \cite{KII-day-night}.  The
exclusion comes from the comparison of the six $\cos\delta_{sun}$ bins,
and is independent of the absolute $^8\mbox{B}$ flux. It is, however, subject
to the uncertainties from the Earth density profile in the MSW calculations.

In the combined fit including the Kamiokande day-night data, we treat
the six
$\cos \delta_{sun}$ data as independent data points, along with the
Kamiokande III, Homestake, and the gallium results.
The normalized day-night data were taken
from  Ref.~\cite{KII-day-night}.  They were scaled to give the average
Kamiokande II rate; also, we included   the overall systematic uncertainty
(15\%) from the energy calibration, the angular resolution,
and the event selection.   We checked for consistency by
combining the six bins and reproducing the time-averaged Kamiokande II data
($0.47 \pm 0.08$ relative to the Bahcall-Pinsonneault prediction).
In the $\chi^2$ fits, we calculate a 9$\times$9 error matrix
for each $\Delta m^2$ and $\sin^22\theta$.
The Kamiokande II systematic uncertainty  is correlated
among the angular bins
\footnote{
We ignore possible correlations of the systematic uncertainty
in the Kamiokande II and III data.
};
the flux and cross section uncertainties are also included and correlated
between the nine constraints.

The result of the combined fit including the Kamiokande II day-night data
is shown in Fig.~\ref{fig_dn}.  The absence of
enhancement in the nighttime Kamiokande
data constrains the large-angle solution, leaving only a small allowed region
at 90\% C.L.  The nonadiabatic solution is unchanged by the Earth effect
and is still preferred.  The minimum $\chi^2$ values
are 3.1 and 6.8
for the nonadiabatic and large-angle regions, respectively; both
 are nominally good fits for 7 degrees of freedom (= 9 data -- 2 parameters),
although the large-angle solution gives a poor fit without the Kamiokande II
day-night data \cite{BHKL}.  The very small $\chi^2$ contribution
(3.0) from the six Kamiokande II angular bins is responsible for the
improvement.  The best fit parameters with $\chi^2$ value are
summarized in Table~\ref{best-fit}.  The allowed regions for 90, 95, and
99\% C.L. are shown in Fig.~\ref{fig_CLs}.

The uncertainties in the Earth density profile are not included in the
above calculations.  To examine the effect, we varied the core density and
the core-mantle boundary by $\pm$10\%.  No difference was observed for the
combined fit in Fig.\ \ref{fig_dn}.

It is pleasure to thank Eugene Beier, Sidney Bludman, Ed Frank,  and
Bill Frati for useful discussions.  This work was supported by
the Department of Energy Contract DE-AC02-76-ERO-3071.

\newpage
% Here comes the reference ****************************************************

%                        * * * REFERENCE * * *

\newpage
% Here comes the tables *******************************************************

\begin{table}[p]
\caption{
%
%                                  TABLE I
%
The time-averaged solar neutrino rate relative to the standard solar model
predictions by Bahcall and Pinsonneault \protect\cite{Bahcall-Pinsonneault}.
The Homestake result \protect\cite{Homestake} is updated from 0.26$\pm$0.04
used in Ref.\ \protect\cite{BHKL}.  Also listed are the latitudes of the
solar neutrino detectors.
}
\label{experiments}
\vspace{1.0ex}
\begin{tabular}{l  c c c c c}
%
%------------------------------------------------------------------------------
%------------------------------------------------------------------------------
              &  Homestake & Kam II & Kam III & SAGE & GALLEX               \\
\hline%------------------------------------------------------------------------
 Rate (SSM)& 0.28$\pm$0.03 & 0.47$\pm$0.08 & 0.56$\pm$0.09
                             & 0.44$^{+0.17}_{-0.21}$  & 0.63$\pm$0.16      \\
 & &\multicolumn{2}{c}{0.50$\pm$0.07} & \multicolumn{2}{c}{0.54$\pm$0.11}   \\
\hline %----------------------------------------------------------------------
 Latitude  & 44.3$^\circ$ N & \multicolumn{2}{c}{36.4$^\circ$ N}
				      & 43.5$^\circ$ N & 42.4$^\circ$ N     \\
%------------------------------------------------------------------------------
%------------------------------------------------------------------------------
%
\end{tabular}
\end{table}

%                     *               *               *

\begin{table}[p]
\caption{
%
%                                  TABLE II
%
The best fit parameters and $\chi^2$ values for the two  regions allowed by
the combined MSW fit including the  Earth effect.  The Kamiokande II
day-night data of the six angular bins are incorporated in the fit.  The
time-averaged data are used for Homestake, Kamiokande III, SAGE, and GALLEX.
}
\label{best-fit}
\vspace{1.0ex}
\begin{tabular}{l  c c }
%
%------------------------------------------------------------------------------
%------------------------------------------------------------------------------
                       &  Non-adiabatic        & Large-angle                \\
\hline %-----------------------------------------------------------------------
 $\sin^22\theta$       & $7.8\times 10^{-3}$   & 0.68                       \\
 $\Delta m^2$ (eV$^2$) &  $0.55\times 10^{-5}$ & $1.0\times 10^{-5}$        \\
 $\chi^2 $ (7 $d.o.f.$)&   3.1                &  6.8                        \\
%------------------------------------------------------------------------------
%------------------------------------------------------------------------------
%
\end{tabular}
\end{table}

\newpage
% Here comes the figures ******************************************************

\begin{figure}[p]
%\postscript{fig1.ps}{0.80}                                     %**epsf**
\vspace{2ex}
\caption{
%
%                                  FIGURE 1
%
The allowed regions including the Earth effect of the Kamiokande, Homestake,
and gallium experiments, and the combined fit at 90\% C.L.  The time-averaged
data are used.  The Bahcall-Pinsonneault model and uncertainties are assumed
in the calculations.  Also shown  is the large-angle solution when the Earth
effect is ignored.
}
\label{fig_ave}
\end{figure}

%                     *               *               *

\begin{figure}[p]
%\postscript{fig2.ps}{0.80}                                    %**epsf**
\vspace{2ex}
\caption{
%
%                                 FIGURE 2
%
The allowed regions including the Earth effect of the Kamiokande, Homestake,
and gallium experiments, and the combined fit at 90\% C.L.   Kamiokande II
day-night data with six data points are included.  Also shown is the excluded
region (grey) using only the Kamiokande II day-night data.  This exclusion
comes from the comparison of the six angular bins and is independent of the
$^8$B flux uncertainty.
}
\label{fig_dn}
\end{figure}

%                     *               *               *

\begin{figure}[p]
%\postscript{fig3.ps}{0.80}                                    %**epsf**
\vspace{2ex}
\caption{
%
%                                FIGURE 3
%
The allowed regions of the combined fit including the Kamiokande II day-night
data at 90\%, 95\%, and 99\% C.L.
}
\label{fig_CLs}
\end{figure}

% The End ********************************************************************
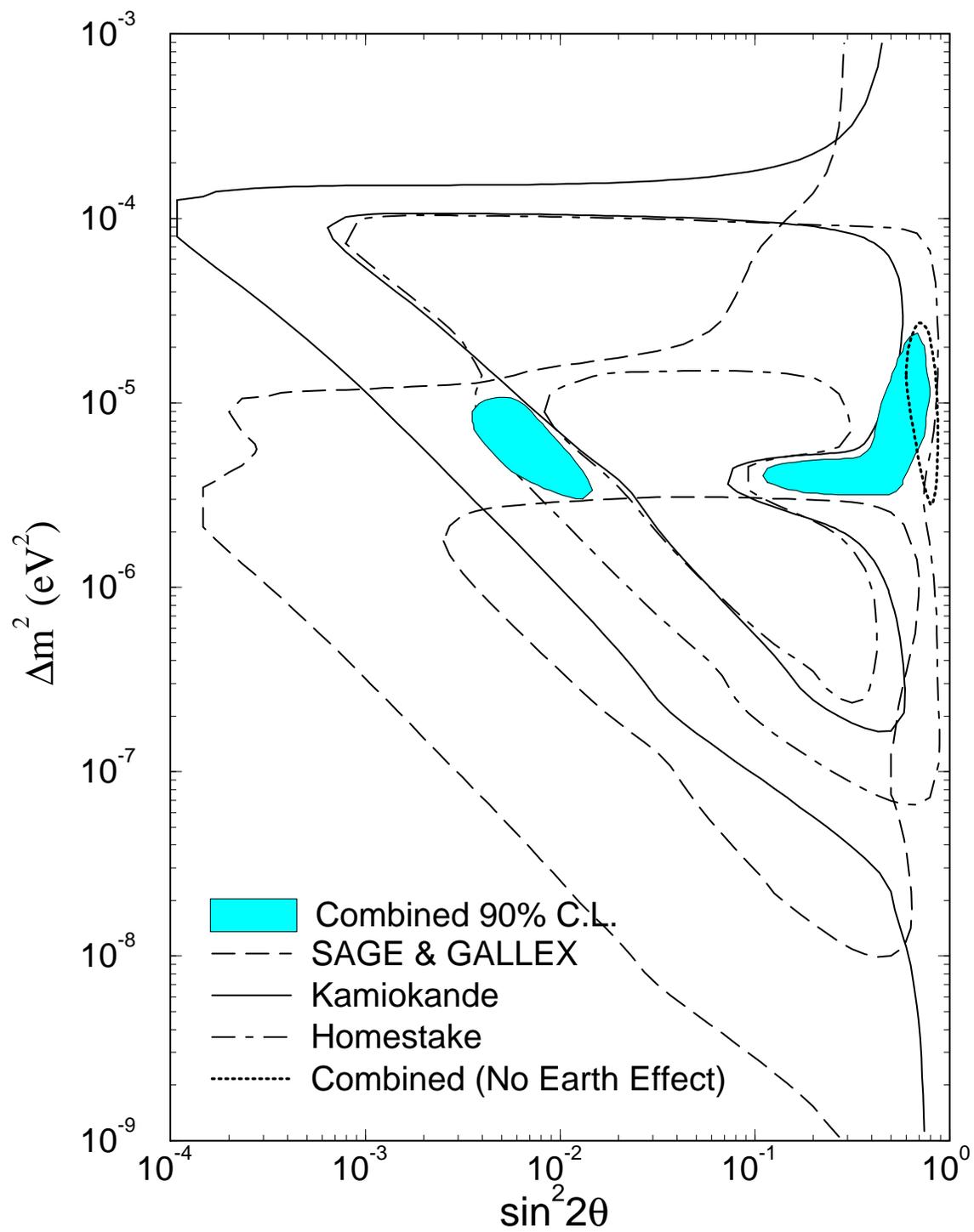
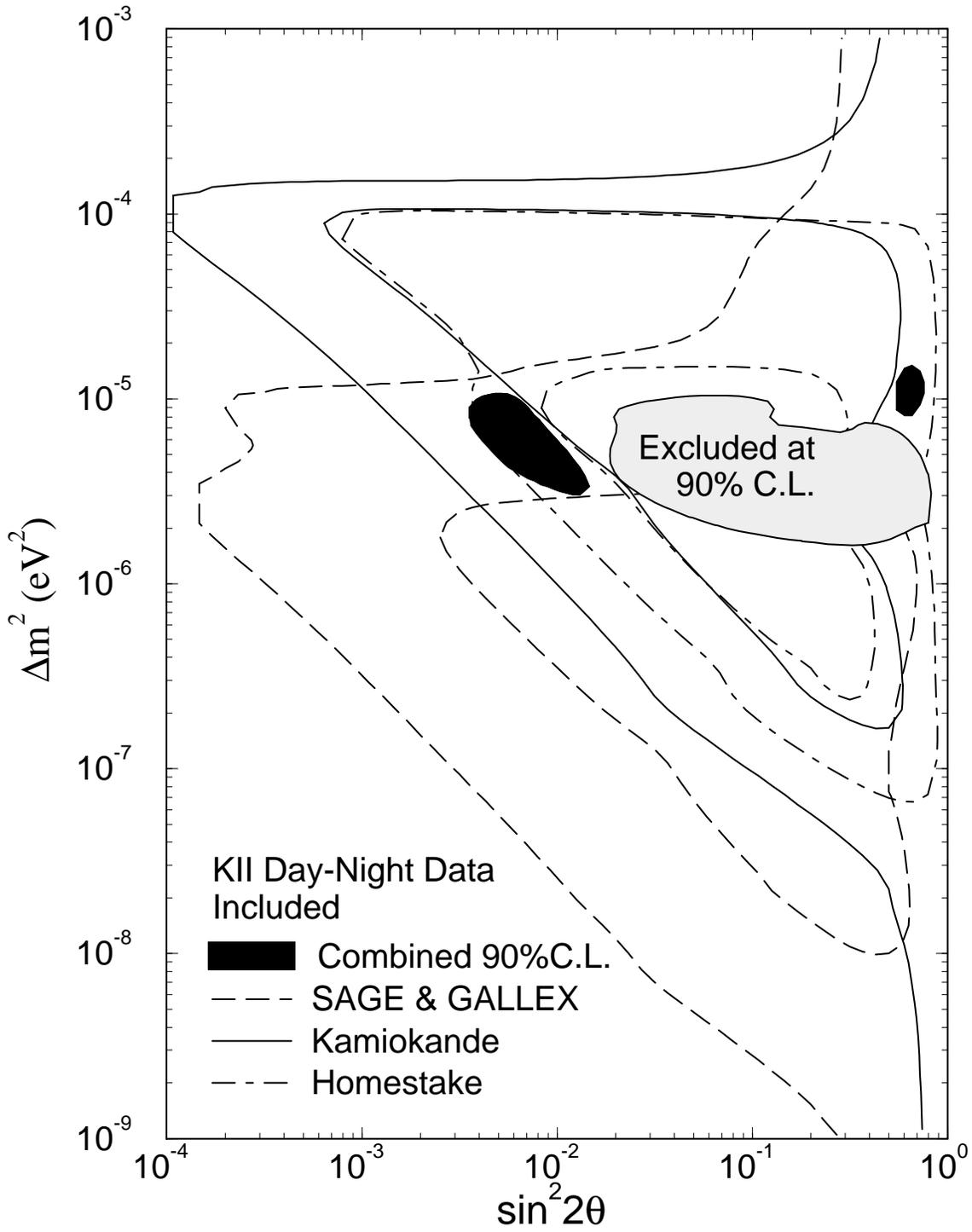
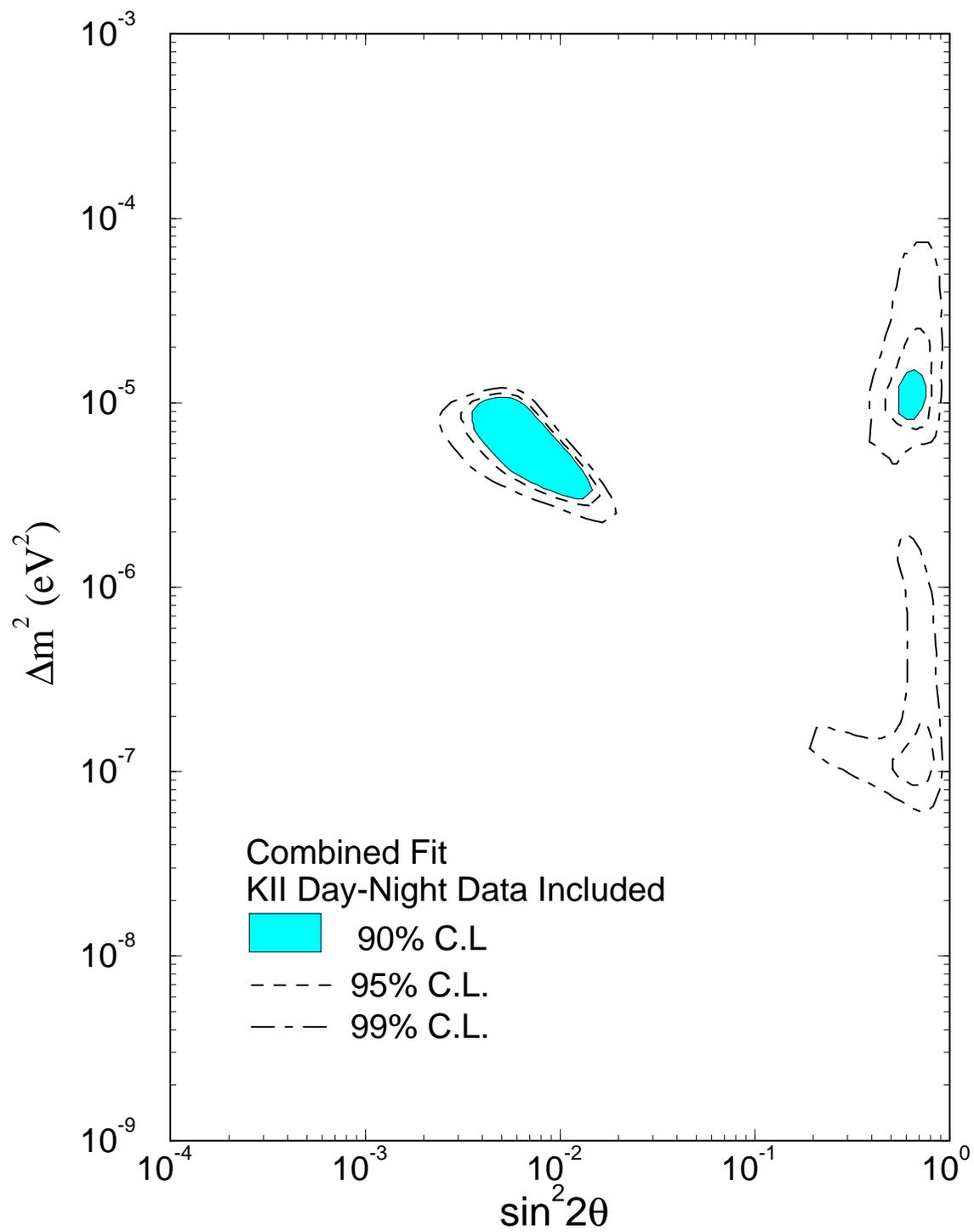
\end{document}